# Metrics-Based Spreadsheet Visualization
## *Support for Focused Maintenance*

*Karin Hodnigg, Roland T. Mittermeir* [1]

[1] *Institut für Informatik-Systeme*
*Alpen-Adria-Universität Klagenfurt*
*AUSTRIA*
*Karin.Hodnigg@uni-klu.ac.at*, *Roland.Mittermeir@uni-klu.ac.at*

**ABSTRACT**

*Legacy spreadsheets are both, an asset, and an enduring problem concerning spreadsheets in business. To make spreadsheets stay alive and remain correct, comprehension of a given spreadsheet is highly important. Visualization techniques should ease the complex and mind-blowing challenges of finding structures in a huge set of spreadsheet cells for building an adequate mental model of spreadsheet programs.*

*Since spreadsheet programs are as diverse as the purpose they are serving and as inhomogeneous as their programmers, to find an appropriate representation or visualization technique for every spreadsheet program seems futile. We thus propose different visualization and representation methods that may ease spreadsheet comprehension but should not be applied with all kind of spreadsheet programs. Therefore, this paper proposes to use (complexity) measures as indicators for proper visualization.*

## 1 INTRODUCTION

Software that is used is subject to continuous modification and change [Lehman, 1980]. This "law" apparently applies independent of the representation of this software; specifically it is independent of whether it is expressed in a conventional procedural or object-oriented programming language or in a fourth-generation spreadsheet system. Hence, as with general software, maintenance of spreadsheets or spreadsheet programs requires that the person performing maintenance comprehends not only the details to be changed but also to an appropriate extent the whole system where this change is to be applied. However, the comprehension issue with spreadsheets involves different concepts and capabilities than needed when attempting to comprehend conventional programs.

Before concentrating on these differences, some terminological clarifications: We use the term "spreadsheet program" for the artefact produced by an application expert (spreadsheet programmer). The term "spreadsheet system" is used to refer to the implementation environment e.g. *Excel*, *Calc*, *Gnumeric* etc.

### 1.1 Structure of the Paper

The paper starts out by focussing on the very nature of spreadsheet programs in contrast to conventional programs and the ensuing differences for spreadsheet reverse engineering. Based on these considerations, different maintenance issues are addressed. They are contrasted with various visualization approaches proposed so far.

This stipulates the question: Which one of these visualization approaches is best. We conclude that there is not a single answer to this question. Contrary, the answer to this question depends on the nature of the spreadsheet programme, the programmers qualification and the specific maintenance task. A professional spreadsheet programmer might know the answer for her or his problem right away. Casual spreadsheet programmers might stumble with this question though. For them, we propose a set of metrics that might help choosing the most suitable visualization for the problem at hand.

**1.2 Legacy Spreadsheets and Spreadsheet Reverse Engineering**

Legacy spreadsheets are – similar to common legacy software – programs that are "*vital*" to the organization [Bennet, Rajlich, 2002], but undergo no further development. Missing (tacit) knowledge may lead to a "don't touch it"-policy. On the other hand, most spreadsheet programmers are end users and as such no software engineering experts or programmers. Thus, they rarely comply with a defined development process. Their artefacts are weakly tested and adequately documented.

It is also important to see that spreadsheet systems provide no explicit distinction between *programming* and *running the program* and that from the perspective of the developer or the user, there is no clear distinction between data and code. Consequently, it is quite easy to mistype something and change code instead of data. Therefore, spreadsheet programs age even more rapidly than compiled software artefacts. Although spreadsheet applications offer security mechanisms to avoid accidental change of formulas or misuse, it requires a bit of expertise to use these mechanisms.

Spreadsheet programs are not deliberately degraded from working to legacy spreadsheet. They succumb to their own (rapid) aging process due to the programming environment and its characteristics, user experience, organizational frameworks etc. In [Bennett, Rajlich, 2000], the authors also state that for "*... many organizations, the data is the strategic asset, rather than the code*". This is especially true for spreadsheets where the data is an *inherent* part of the program and thus, spreadsheet comprehension should be considered a high priority task (whether data or computation is dominant).

Spreadsheet Reverse Engineering may be defined as the analysis of a spreadsheet program to identify data and their interdependencies including the documentation, to map the program onto a (highly) abstract representation. With the objective of design recovery, the first reverse engineering task is the translation of a (complex) spreadsheet program into an abstract model. This model allows identifying the structures and potential hot spots for targeted maintenance and streamlined evolution. Therefore, we focus on spreadsheet comprehension as precondition for any re-engineering.

**2 SPREADSHEET COMPREHENSION**

Spreadsheet programs are usually seen as relatively simple arrangements of a few formula or data cells. This might be true for the products of students and novices. But application experts do develop spreadsheet programs of high complexity which soon have a number of cells filled by data or formulae going well above 1000. Hence, any attempt aiming at comprehending the sheet by simply perusing all cells (or only all formula cells) is bound to failure by exhaustion.

Such large spreadsheet programs require at least a definitive reading strategy. With conventional programs several reading strategies have been developed [Biffl, 2001]. They more or less follow either linear textual order, the control flow, or some specific data flow through the program. With spreadsheet programs such evident hints are not available. On the contrary, people quite often start perusing the spreadsheet program at the value layer (c.f. Fig. 1) where the sheet manifests itself only as a (partial) two-dimensional arrangement of (result-)values. Only the cell pointed at by the cursor is displayed in terms of its semantic content, either data or the formula used for computing the result shown on the value layer.

Thus, as shown in Fig. 1, every spreadsheet programmer is confronted with three layers: the *value layer*, where the resulting values (including a potential formatted representation) of every cell are presented, the underlying *formula layer*, which shows the structure of the formulas, and the layout independent *data dependency graph*. The formula view offered by the different spreadsheet systems is still layout dependent, whereas the underlying data dependency graph is detached from the geometrical layout. It is displayed only by tools that can cooperate with specific spreadsheet systems.

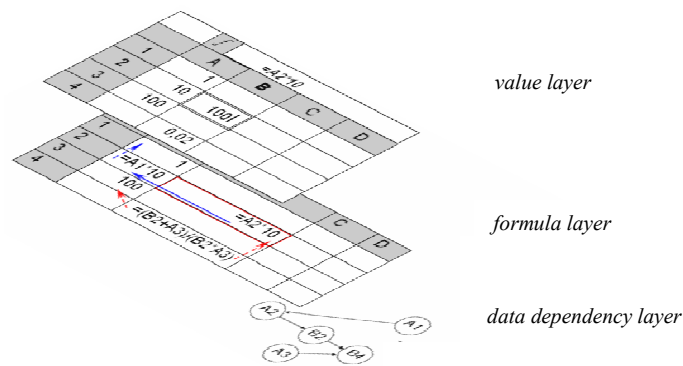

**Figure 1. Three layers of spreadsheet programs**

The task of understanding a given spreadsheet program involves different levels of abstraction that result from the specific nature of the arrangement of formulas.
- The level of individual formulas: Here, questions such as where are the arguments of this formula computed and where is the result computed by this formula is used further on (except for the display on the value layer).
- The two-dimensional arrangement of formulae on a worksheet. Here the question to which extent the sheet can be separated into relatively independent sub-areas and whether there is some pattern in the flow of information from data cells via formula cells to those formula cells that yield those results which are finally of interest.
- The three-dimensional arrangement of formulae and consequently of data flow throughout a workbook.

One should note, that not all spreadsheet programs utilize the third dimension, i.e. links between different sheets contained in a workbook. Further links, e.g. links between workbooks, can be seen as further dimensions. However, there are enough examples where this third dimension is used in a very appropriate manner. One might think about monthly result sheets and a thirteenth sheet accepting summary values from each month and performing the appropriate end-of-the-year computations. Alternatively one might think of a strategy sheet, computing corporate values which are handed down to departmental sheets in such a way that they can be connected with departmental data on the individual background sheets. Possibly, the results of these computations are afterwards again handed forward to the corporate sheet (or handed further backward to another corporate sheet).

Thus, comprehension of a workbook is non-trivial as there are several factors that aggravate its comprehension. Foremost among them is the inherent nature of spreadsheet applications partly hiding data dependencies.

To understand a given (complex) spreadsheet, a spreadsheet programmer/maintainer usually builds the mental model incrementally. The data dependencies often form a huge graph. Spreadsheet programmers usually find a path through this graph – either by clicking into a cell and discover its predecessors and successors (with tracing tools) and moving along a tempting/interesting path, moving from cell to cell. Alternatively, one may try understanding the functionality by using additional information in cells or layout. The spreadsheet comprehension process is generally assumed to be opportunistic, though we are missing confirmative studies.

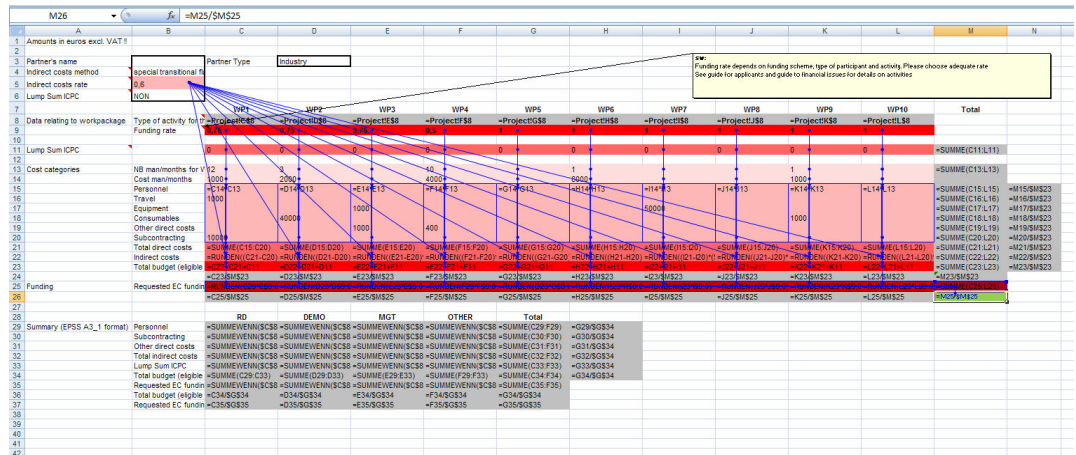

**Figure 2. Analysis of dependencies of cell M26**

Although some spreadsheet systems offer the possibility to highlight values and formulas in different colors (e.g. *OpenOffice Calc*) and/or to incrementally build the data dependencies with blue arrows (e.g. *Microsoft Excel's tracer*) as shown in Fig. 2, these visualizations are still only rudimentary (and local!) and do not provide an overview of the spreadsheet program's structure. Even Excel tracer meets its limits rather soon; if a cell reference exceeds worksheet boundaries, the reference cannot be properly visualized.

**2.1 Spreadsheet Comprehension Impediments**

Thus, spreadsheet comprehension is impeded by different factors: first, the inherent properties of the programming environment with its hidden data dependencies hinder a structural grasp. The fact that portions of a worksheet might be accessible only via scrolling is a hurdle, but might not be significant. The fact that information flows over worksheets within a workbook is however a clear impediment to comprehension efforts.

The informal and unconstrained mix of input data and actual spreadsheet model exacerbates maintenance (as discussed above). The functional formula language provides some quite complex formula constructs (e.g. `*LOOKUP()`, `INDIRECT()`, `OFFSET()`, statistical and financial formulae that must be understood). Nested expressions and conditionals are also challenging. In this context we point to the fact that in procedural languages, indentation rules are proposed to support comprehension of conditional parts of the program. In spreadsheet programs, expressions are to be written as linear text into a single cell. Highlighting of scopes by displaying parentheses in different color is about the only aid available for visually parsing a complex conditional or nested expression.

**2.2 Spreadsheet Comprehension Facilitation**

The nature of spreadsheet development plays also a special role for spreadsheet comprehension. A typical feature of developing a spreadsheet program is to key in a certain formula and copy it afterwards over a connected area of cells or moving it to some remote portion of the sheet. From the final spreadsheet program one cannot be sure that this development approach has been used. However, the layout and documentation possibilities (building of coherent blocks, discriminating frames, etc.) might help to formulate appropriate heuristics such that copied formulae or even copied regions containing different formulae have some common semantic root.

A mental model, describing the "*person's internal representation*" [Ramalingam, 2004] of the program, is for spreadsheet developers "*almost certainly line-, column- or block-based*" [Mittermeir, Clermont, 2002]. In general, this notion eases the comprehensibility of a spreadsheet program, since layout characteristics help to identify related concepts alleviating spreadsheet comprehension. The underlying programming paradigm is easy to grasp, as discussed in [Hodnigg et al., 2004] and [Hodnigg, 2005].

**3 DIFFERENT VISUALIZATION APPROACHES**

This section provides an overview of selected spreadsheet visualization approaches and proposes a different approach to the data dependency graph. Spreadsheet visualization has to fulfill different requirements since spreadsheet users follow an opportunistic advancement in the comprehension process (top-down and bottom up). This means, different representation levels have to be offered, as well as navigability and interactivity. The integration of spreadsheet visualization must be seamlessly integrated into the spreadsheet system, otherwise repellant user feedback must be expected.

The different representation levels start from a cell level. There, it should be possible to explore the formula. Proceeding to a module level that illustrates e.g., semantic classes [Mittermeir, Clermont, 2002] or data blocks [Clermont 2003b] can help to drastically reduce the number of cells to be perused before comprehending a spreadsheet program. Further, a worksheet and workbook representation should show the overall (geometrical) structure of a spreadsheet program.

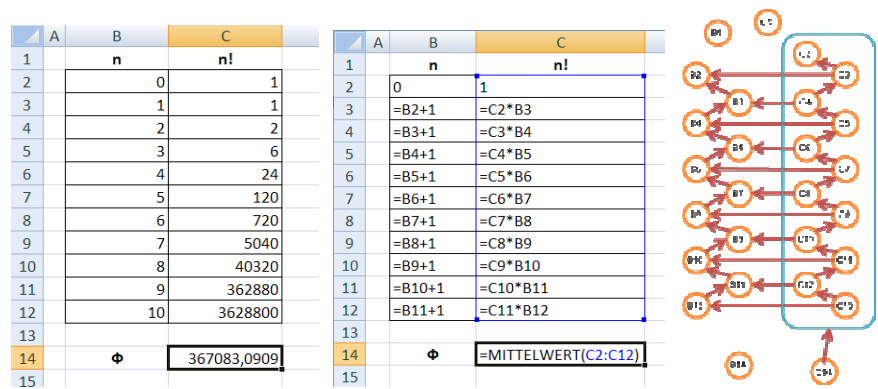

**Figure 3. Three views on a spreadsheet program (factorial list)**

Common representations of spreadsheet programs are data dependency graphs as observed in [Ballinger et al., 2003] and [Ayalew, Mittermeir, 2003]). In [Yoder, Cohn, 2002], the authors define inter-cell dependencies concisely: "*When a cell $c_1$ makes a reference to another cell $c_2$, $c_1$ becomes dependent of $c_2$*". The cell $c_1$ points to its predecessor $c_2$ and changes in $c_2$ propagate and influence the value of cell $c_1$. In a spreadsheet program, a formula reference can point to a range of cells (as shown in Fig. 3), e.g. =SUM(A1:A12). There are differences in the handling of range references as discussed in [Hodnigg et al., 2004], so the incorporation of range references into the spreadsheet dependency graph is required.

It is noticeable that the direction of the vertices in Fig.3 does not correspond to the data "flow" direction, but originates from the visibility order proposed by Mittermeir and Hodnigg in [Hodnigg et al., 2004]: Here, the idea of visibility of cells represented as projection-screen devices (with internal computational and external visualization unit) is introduced to explain the spreadsheet paradigm through visibility but linked dependency. The thus so-called projection-screen model is basis for other educational approaches to explain the spreadsheet paradigm to Austrian high school students, where its use indicated better understanding. Thus, the visualization in Fig. 3 sticks to the idea of visibility in contrast to a data flow graph. Exceeding the given definition of (simple) graphs, hypergraphs are graphs where edges can connect sets of nodes [Gallo, Scutella, 1999], [Thakur, Tripathi, 2005]. Range references can thus be expressed through directed hyperedges. The following definition extends the definition in [Thakur, Tripathi, 2005].

**Definition 3.1 Data Dependency Hypergraph.** A *directed hypergraph* $\mathcal{H}$ is a tuple of *(V, E)* where $V = \{v_1, ..., v_n\}$ is the set of vertices and $E \subseteq \mathcal{P}(V) \times \mathcal{P}(V)$ is the set of hyperedges, such that $e = \{T(e), H(e)\} \in E, T(e) \neq \emptyset, H(e) \neq \emptyset$ and $H(e) \cap T(e) = \emptyset$.
$H(e)$ is the "head-set", $T(e)$ the "tail-set" of the hyperedge. An *F-hyperedge* is a hyperedge in which for every $e \in E: |T(e)| = 1$, an *F-hypergraph* is a hypergraph $\mathcal{H}$ such that each hyperedge $e \in E$ is an F-hyperedge. If the *F-hypergraph* $\mathcal{H}$ contains no circles, it is acyclic.

Let *h* be a directed hyperedge defined as $h=(\{s_1,s_2, ..., s_n\}, \{d_1,d_2, ..., d_n\})$, where $H(e) = \{s_1,s_2, ..., s_n\}$ is the source (head- set) and the set $T(e) = \{d_1,d_2, ..., d_n\}$ is the destination set (tail-set) of the hyperedge. Head and tail are *disjoint* sets. In a spreadsheet data dependency hypergraph, the cardinality of the source set $|\{s_1,s_2, ..., s_n\}|=1$. Thus, we may simplify the directed hyperedge to a tuple $(s,\{d_1,d_2, ..., d_n\})$. That is, the directed edge has its source in one distinct cell *s* but may reference more than one cell.

The example in Fig. 4 illustrates the definition of a sample F-Hypergraph ({A1, A2, B3, B4, B5, B6}, {$e_1$, $e_2$, $e_3$}) , where $e_1$ and $e_3$ are simple edges and $e_3$ is a directed F-hyperedge from vertex 3 to the vertex set {B4, B5, B6}.

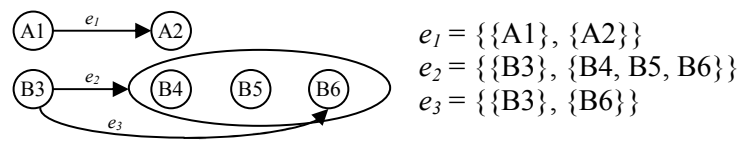

$e_1 = \{\{A1\}, \{A2\}\}$
$e_2 = \{\{B3\}, \{B4, B5, B6\}\}$
$e_3 = \{\{B3\}, \{B6\}\}$

**Figure 4. A small hypergraph example**

A formula corresponding to the lower part of Fig. 4 might be B3 = sum(B4:B6)+B6. This might not be the most usual way of writing a formula. But the example demonstrates that the receiving node might have multiple links into its source set. In general, the representation of a spreadsheet as a data dependency hypergraph will be quite complex. Thus, by itself it does not suffice to ease spreadsheet comprehension. No abstraction is given and the spreadsheet programmer has to explore the whole structure incrementally. With abstractions such as mentioned in [Mittermeir, Clermont 2002] and in [Clermont 2003b] as well as with the equivalence classes discussed below, further aggregations can be obtained. Thus, the hypergraph will lose its F-property and become a full hypergraph. However, the resulting structure will become already quite abstract and, therefore, further visualization support is called for.

### 3.1 Related Work

Most visualization efforts concentrate on the data dependency graph structure, as discussed in [Ballinger et al., 1999]. The *data dependency flow visualization* should be emphasized: spatial flow direction of references is used to visualize the direction of data dependencies. For this visualization, the spreading factor of formulae should be rather low, as should the number of cells. The *spring view* "*based upon the idea of spring forces to arrange the spatial positions of cells*" [Ballinger et al., 1999] shows the dependency structure of given cells and their interdependencies.

Brath and Peters focus on the data aspect in [Brath, Peters, 2006]. Values of cells are displayed in a 3D manner, so that patterns and ruptures within them can be recognized by domain experts. This visualization approach is promising for data centred spreadsheet programs. Another tempting and promising visualization is described in [Wettel et al., 2007] where a city metaphor is stressed to visualize object-oriented software, but it can be adapted to spreadsheet visualization requirements, if the spreadsheet program is calculation-centred. Other object-oriented visualization approaches may apply, too.

### 3.2 A Scalable Approach to Spreadsheet Visualization

The model visualization approach by Clermont in [Clermont, 2003a] is based on the assumption, that "*spreadsheet programmers have a conceptual model of the spreadsheets they create […] that determines how cells can be grouped into units*". The reconstruction of the spreadsheet model is performed with the aid of three (four, if [Hipfl, 2004] is considered) visualization approaches:
  i. *Logical Areas* are based upon the similarities of cell formulas irrespective to their spatial distribution
  ii. *Semantic Classes* are regularly recurring cell areas with similar neighbourhood
  iii. *Data Modules* are sets of cells that contribute to a given result cell
  iv. *Layout Areas* ([Hipfl, 2004]) assigns sets of cells to given labels using geometrical or semantical information.

Approaches (i.) and (ii.) are discussed in detail in [Mittermeir, Clermont, 2002], [Clermont, 2003b]. Approach (iii.) is presented in [Clermont, 2003a *and* 2005]. With (iv.), the author further develops these approaches by integrating different concepts of semantic and geometric assignments [Hipfl, 2004]. These different visualization approaches are highly customizable and provide audit information at a more sophisticated level than Def. 5.2.6. Nevertheless, the number of logical areas, semantic classes, data modules and layout areas each built with a specialized parameter set passed to the visualization functions provides the user with an overview of the spreadsheet program, therefore, hotspots resulting from design errors can be identified. Evaluative experiments on these approaches have been reported in [Clermont, 2003a].

### 3.3 A 3D Workbook Visualizer and A Dual Spreadsheet View

This approach aims for visualization at a very pragmatic level. One of the main obstacles to spreadsheet comprehension is the worksheet barrier – e.g. for Excel tracer, already for scrolling operations when entering a formula. One could imagine a 3D visualization of stacked worksheets. Since the dependency representation would exceed comprehension goals, the units presented in such a 3D workbook visualizer must be of abstract or aggregative form, such as data modules. Such a visualization must of course, be rotatable, zooming in into different layers must be possible.

- For a change analysis task, this representation is very attractive since change effects can be visualized overcoming worksheet barriers.
- Common patterns of spreadsheet programs, such as e.g., monthly sheets and an aggregation sheet, can be easily identified with this approach.

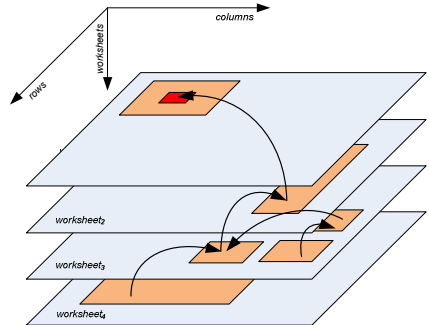

**Figure 5. 3D Worbook Visualizer**

For cells with high fan-out and long calculation chains, this representation would be able to present the actual slice of the spreadsheet program that would be affected by changes. Evidently, it is also possible to make the individual sheets transparent. In this case the adherence of a node to a particular sheet can be highlighted by using different colours for nodes belonging to different worksheets.

Apart from the main focus of this paper, a dual view as spreadsheet visualization should be proposed: Adapted to the user's needs, he/she should be provided with conventional user interface, enhanced with a window, where different visualizations of the actual spreadsheet program can be explored, navigated through and interacted with. Every visualization must stick to the spreadsheet paradigm, not forcing the user to learn representation attributes he/she is not interested in. Spreadsheet Comprehension is sufficiently complex.

**4 SOME FORMAL DEFINITIONS**

Spreadsheet programs can be understood as a set of cells that are connected through dependencies. Typically, a spreadsheet program does not include circular references, although they are permitted. These circles are – in most cases (except for scientific purposes) – design errors, so most spreadsheet systems warn the user before such a reference can be accepted. Hence, we focus on acyclic spreadsheet programs in the following chapters, circular references will be discussed where necessary, though. In a spreadsheet program, the vertices are cells. Cells may have different roles in a spreadsheet program, e.g. a cell containing a value may be input data serving the actual calculation or a label (documentation). We thus distinguish between different types of cells. The following definitions are based on the formal definitions in [Clermont, 2003a] and [Sajaniemi, 2000]

**Definition 4.1 Cell.** A cell is a triple $c = (ca : CellAddress, v:V, f:(...) \to V)$, where the cell address $ca$ is a ordered tuple $ca = (ca_1, ..., ca_n)$, $ca_i \in \mathbb{N}$ and the value $v \in V = \mathbb{Q} \cup Strings \cup \{Error\} \cup \{Undefined\}$. $f$ is the cell formula $f:(cref^+) \to V$, where $cref: (c_{src}, \{c_1, ..., c_r\}) \to V$ is the cell-referencing function. The *set of all cells* $c$ is $C$. The set $\{c_i = (ca, v, f), c_i \in C \mid v \in \{Undefined\} \land f = \emptyset\}$ is the *set of empty cells* $C_e$.

The cell address $ca$ defines the geometrical position of the cell, the value $v$ is the value displayed as a result of the evaluation of the formula $f$. If the cell contains a value, but no formula, this value is displayed, if c specifies a formula, the result value is displayed: $c = ((ca_1, \ldots, ca_n), v, f)$, $f \neq \emptyset, v = eval(f)$. If an empty cell $c_e \in C_e$ is referenced with $cref(c, c_e) = 0 \; \forall c \in C$, the assumed value is 0. The function $cref$ follows the data dependencies, formed by the cell references of the formula language.

```
Formula ::=              EQ FormulaExpression | ArrayFormula
FormulaExpression ::=    Expression | Expression Operator Expression
Expression ::=           LPAREN FormulaExpression RPAREN
                         | ReferencePrefix | FunctionCall | NUM
                         | ArrayConst | STRING | STRING1 | Text
LogicalExpression ::=    FormulaExpression LogOp FormulaExpression
                         | FormulaExpression
…
Operator ::=             AMPAND | MathOp
MathOp ::=               PLUS | MINUS | MULT | DIV | CNTRL
LogOp ::=                GT | LT | EQ | NEQ | GTEQ | LTEQ
…
FunctionCall ::=         Function LPAREN ArgList RPAREN
                         | IF LPAREN LogicalExpression SCOLON FormulaExpression RPAREN
                         | IF LPAREN LogicalExpression SCOLON FormulaExpression SCOLON
                         FormulaExpression RPAREN
                         | LOGFUNC LPAREN LogArgList RPAREN
                         | NOT LPAREN LogicalExpression RPAREN
…
```

**Figure 6. Excerpt from the Spreadsheet Language Definition**

Fig. 6 shows a simple BNF grammar for the formula language. Each formula starts with the equality sign, followed by a formula expression, potentially combined with mathematical operators. A formula expression can be a function call of built-in or user-defined functions.

**Definition 4.2 Spreadsheet program.** A *spreadsheet program* is an acyclic F-Hypergraph $S = (V, E)$ with the set of vertices $V = C \setminus C_e$ and the set of hyperedges $E = \mathcal{P}(V) \times \mathcal{P}(V)$. A hyperedge is formulated by references: $e \in E$ if there exists a dependency between $e_{src}$ and the set $T(e) = \{c_1, \ldots, c_n\}$ with $c_{src} \notin T(e)$ and $\{e = (c_{src}, T(e)) \mid \exists \, cref(c_{src}, T(e))\}$.

**Definition 4.3 Degree of a Cell.** Let $G = (V, E)$ be a graph. Then the degree $deg_G(v)$ of a vertex $v$ is the number of incident cells. A distinction is drawn between the indegree (incoming) $deg_G^+(v) = |\{c \in T(e) \mid (v, T(e)) \in E\}|$ and outdegree (outgoing edges) $deg_G^-(v) = |\{c \in H(e) \mid (H(e), v) \in E\}|$.

**Definition 4.4 Cell types.** Let $S = (V, E)$ be a spreadsheet program. Then, $C_v$ is the *set of value cells* set with. $C_v = \{c_i = (ca, v, f), c_i \in C \mid v \in V \wedge f = \emptyset\}$. Every $\forall c \in C_v$ contain user-defined values. This set can be split into data and label cells.
Label cells are isolated nodes. The *set of label cells* $C_l$ then is $\{c \in C_v \mid deg_G^+ = 0 \wedge deg_G^-(v) = 0\}$.
Data cells are cells that serve as input for calculations. The *set of data cells* $C_d$ can be defined as $\{c \in C_v \mid deg_G^+ > 0 \wedge deg_G^-(v) = 0\}$. The property $deg_G^-(v) = 0$ is immanently true for both label and data sets, since there are no references $\nexists \, e \in E \; \forall c \in C_v : e = (H(e), v)$, as $f = \emptyset$.
The *set of formula cells* $C_f$ can be defined as $\{c = ((ca_1, \ldots, ca_n), v, f), c \in C \setminus C_v \mid f \neq \emptyset\}$.

A label cell is a cell that is not referenced, but may serve as documentation, unit description or comment. The definition of labels differs somewhat from the label definition found in [Hipfl, 2004], since only constant labels are considered. One could extend $C_l$ to $C_l'$, in incorporating the defined labels (e.g. labels based on simple computations "=R[1]C+1" such as a sequence of days), ibid.

**Definition 4.5 Coupling Matrix.** Let $S = (V, E)$ be a spreadsheet program. Then, $S$ can be described by a matrix. There, $m(c_i, c_j)$ denotes the number of edges leading from $c_i$ to $c_j$. A matrix of dimension ($n \times n$) is called *coupling matrix A* (analogous to the adjacency matrix) as $A = (a_{ij}) = m(c_i, c_j)$.

## 5 SPREADSHEET (COMPLEXITY) MEASURES

In [Mathias et al., 1999], the authors discuss the impact of size and complexity on software comprehension and comprehension studies. They claim that "*the underlying nature of the software being examined by the programmers […] is a key element […]*". Although their study focuses on object-oriented and procedural programming languages, the results could equally well be transferred to spreadsheet programming and spreadsheet comprehension. In [Storey, 2005], the author states that the "*size of the program and other program measures will influence which view is the preferred one to show a programmer browsing the code […]*". This is especially true for end-users, who prefer representations that do not deviate from the original code/program. Thus, a classification of spreadsheet programs has to forego different visualization methods. We have to analyze the complexity, size and structure of spreadsheet programs.

The classification of spreadsheet programs may be done in various ways. Since this paper is foremost interested in visualizations that ease program comprehension, we discuss only a few possible complexity measures. We distinguish between three types of complexity: (a) some general indicators – such as the size of the spreadsheet program, the number of (different) formulas, (b) layout and design complexity and (c) formula complexity (how "twisted" is a formula?) in the face of visualization possibilities. Related efforts can be found in [Bregar, 2004], where the authors mainly focuses on formula and reference structure and complexity. Some of the following complexity measures correspond to his work, but many are newly defined. Especially in [Allen, 2002], different metrics of graph abstractions of software are discussed. [Munson et al., 1992] depicts dynamic metrics.

### 5.1 General indicators

The size of a program always influences the comprehension efforts as discussed in [Storey, 2005], [Mathias et al., 1999], so primarily, a size metric of a spreadsheet program analogously to the Lines of Code (LOC) metric has to be defined. There are discussions whether such a measure as single point of information suffices any complexity estimation [Zuse, 1993], [Shepperd et al., 1997] and even worse, there are different definitions of the LOC metric, but it is agreed that it serves as an important initial information. For spreadsheet programs, such a LOC metric could be represented by different values, e.g. the number of non-empty cells, the number of formulae or the number of distinct formulae.

**Definition 5.1.1 Size of a Spreadsheet Program.** Let $S = (V, E)$ be a spreadsheet program. Def. 2.4 partitioned the set of vertices of the hypergraph $S = (V, E)$ into different sets such that $V = C_l \cup C_d \cup C_f$.
The *size* $sz_v(S)$ is defined by the number of vertices $sz_v(S) = |V| = |C_l| + |C_d| + |C_f|$.
The *enmeshment degree* $ed_e(S)$ is then specified by the number of edges $sz_e(S) = |E|$.
Furthermore, we distinguish between the size of the label set $sz_l(S) = |C_l|$, the data set $sz_d$ and the formula set $sz_d$, as well as the number of hyperedges $ed_h(S)$ and simple edges $ed_s(S)$.

$$sz_l(S) = |C_l| \qquad sz_d(S) = |C_d| \qquad sz_f(S) = |C_d|$$
$$ed_s(S) = |\{e = (H(e), T(e)) \mid |T(e)| = 1\}|, \qquad ed_h(S) = |\{e = (H(e), T(e)) \mid |T(e)| = 1\}|$$

These definitions lend themselves already to a number of interpretations hinting on the very nature of the spreadsheet program one is investigating. E.g., the number of labels $sz_l(S) = |C_l|$ can be understood as a "number of comments" metric. Since label cells do not partake in computations, they can contain useful information as units, structure descriptions. On the other hand, if $sz_l(S) = n$ and $sz_f(S) = 0$, one can confidently assume, that the spreadsheet system was used only for layout purposes, i.e., the sheet one is looking at is actually not a program.

This leads to the question, which share of the spreadsheet program are data, which labels and which actual calculation cells. The ratios $\frac{sz_l(S)}{sz_v(S)} = r_l, \frac{sz_d(S)}{sz_v(S)} = r_d, \frac{sz_f(S)}{sz_v(S)} = r_f$ indicate an additional information ratio, data centeredness, and computational share of a given spreadsheet program $S$. Although a high ratio $r_f$ suggests a high computational complexity, this may not be true for some spreadsheet programs as some of the formulae might result from copy operations. Hence one might be rather interested in the number of distinct formulae or of the relationship $\frac{r_d}{r_f}$ or its inverse, indicating whether the program is rather data centered or formula centered.

Copy equivalence classes defined in [Mittermeir, Clermont, 2002] are used to define the number of distinct formulae. There, Mittermeir et al. defined different forms of equivalence classes of cells, depending on the structure of the formulae, structure of their references etc. Definition 3.1.2 is taken from [Clermont, 2003a], pages 126 ff. and adopted.

**Definition 5.1.2 Equivalence Classes.** Let $S = (V, E)$ be a spreadsheet program. Two cells $c_1 = (ca_1{:}CA, v_1, f_1)$ and $c_2 = (ca_2{:}CA, v_2, f_2)$ are *copy equivalent*, if their formulae are identical. $c_1 \equiv_c c_2 \Leftrightarrow \in f_1 = f_2$. The copy equivalence class $\langle c_1 \rangle_c$ contains all cells that are copy equivalent to $c_1$.

There are other forms of equivalence classes such as logical or structural equivalence. Two cells are *logically equivalent*, if their references differ only in constant values and absolute references, they are *structurally equivalent*, if their formulae contain the same operations in the same order.

**Definition 5.1.3 Number of distinct formulae.** Let $S = (V, E)$ be a spreadsheet program and a given partition of the formula set $C_f = \langle c_1 \rangle_c \cup \langle c_2 \rangle_c \cup ... \cup \langle c_g \rangle_c$, where $c_i \neq c_j \forall i \neq j$. Then $g$ is the number of distinct (copy equivalent) formulae.

The partitions of $C_f$ defined above indicate how "diverse" the formula landscape is, that is, how many different (in the above defined manner) formulas have been used. Analogously, for a logical partition of the formula set $C_f = \langle c_1 \rangle_l \cup \langle c_2 \rangle_l \cup ... \cup \langle c_h \rangle_l$ or structural partition $C_f = \langle c_1 \rangle_s \cup \langle c_2 \rangle_s \cup ... \cup \langle c_k \rangle_s$, $h$ is the number of distinct (logical equivalent) and $k$ is the number of distinct (structural equivalent) formulae. On the other hand, one may be interested in the number of data sources and data sinks.

**Definition 5.1.4 Data Sources and Data Sinks.** Let $S = (V, E)$ be a spreadsheet program. The set $C_v$ is the set of value cells, $C_f$ is the set of formula cells. The set of *data sources* $C_d$ has already been defined in (Def. 2.4) as $\{c \in C_v | deg_G^+ > 0 \wedge deg_G^-(v) = 0\}$. More precisely, $C_{src} = C_d \setminus C_l'$. (label blocks are omitted). A *data sink* is defined as a formula cell whose value is not referenced. The set of data sinks $C_{snk}$ then is $\{c \in C_f | deg_G^+ = 0 \wedge deg_G^-(v) \geq 0\}$.

The number of data sources is an overall fan-in metric of $S$ with $sz_{fanin}(S) = |C_{src}|$, analogus $sz_{fanout}(S) = |C_{snk}|$ is an overall fan-out metric.

An induced sub-graph $S_U = (V_U, E_U)$ with $V_U \subseteq V, E_U \subseteq E$ and $E_U$ contains exactly the same edges from $V$ that connect nodes from $V_U$. The metrics defined above can also be used with the spreadsheet program's $S_U$ parts, which may be of special interest. Another criterion for defining sub-graphs might be to look for cells that are displayed in diagrams and charts produced out of the spreadsheet program. Irrespective of whether they are sinks according to the definition given above or not, they represent apparently data relevant for the user.

**5.2 Formula Complexity**

Formula complexity has a huge impact on spreadsheet comprehension. Though the user's individual perception of a formula complexity cannot be estimated, there are objective complexity attributes that can be measured. The longest distance to transitively referenced cell indicates the calculation includes a lot of intermediate steps and is therefore fragile, since any change in the calculation chain leading to the result affects the cell value. The longest distance to a dependent cell, on the other hand, implies that changes in the affected cell may have effects somewhere else on the sheet where they are not expected. Before addressing the formula complexity impact, let us define the basic metric selection.

**Definition 5.2.2 Spreading factor.** Let $c_k \in C_f$ be a formula cell of a given spreadsheet, with $c = (ca:CA, v:V, f:(...) \rightarrow V)$ and $ca = (x_{src1}, ..., x_{srcn})$. There are $r$ references to $r$ cells given in the formula $f$: $E_{c_k} = \{(x_{11}, ..., x_{1n}), (x_{21}, ..., x_{2n}), ..., (x_{r1}, ..., x_{rn})\}$ with $E_{c_k} \subseteq E$. Then, the spreading factor of $c_k$ is the following tuple

$$cc_{spreading}(c_k) = \begin{pmatrix} (min(x_{11},...,x_{r1},x_{src1}), min(x_{11},...,x_{r1},x_{src2}), ... min(x_{11},...,x_{r1},x_{srcn})), \\ (max(x_{11},...,x_{r1},x_{src1}), max(x_{11},...,x_{r1},x_{src2}),, max(x_{11},...,x_{r1},x_{srcn})) \end{pmatrix}$$

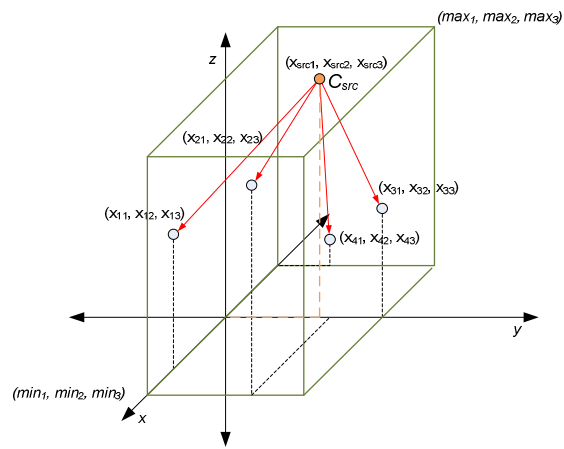

**Figure 7. Geometrical complexity of a formula**

Fig. 7 illustrates the spreading factor of a cell $c_{src}$ with references to four different cells. The dimensions x and y indicate the row and the column of the spreadsheet cell, the third dimension indicates the worksheet the cell is in. A high spreading factor in two dimensions indicates that the references of the cell spread all over the worksheet. A high spreading factor in the third dimension indicates that references are somewhere else in another workbook and thus have to be treated carefully.

**Definition 5.2.1 Calculation Chains.** Let $S = (V, E)$ be a spreadsheet program. Then, $p(c_i, c_j)$ denotes the maximum number of edges leading in an unbroken path from $c_i$ to $c_j$ with $c_i, c_j \in V$. For a given formula cell $c_k \in C_f$ the longest calculation chain is defined as $cc_{pathRef}(c_k) = \max (p(c_k, c_j)),\ c_j \in V$ for referenced cells, $cc_{pathDep}(c_k) = \max (p(c_j, c_k)),\ c_j \in V$ for depedant cells.

For a given cell $c_k \in C_f$ and a spreadsheet program $S = (V, E)$, the set $\{c_j | (c_k, c_j)^+ \in E\}$ describes the **visibility** and $\{c_j | (c_j, c_k)^+ \in E\}$ the **scope** of $c_k$ [Hodnigg et al., 2004].

**Definition 5.2.3 Fan-in and Fan-Out of a cell.** Let $c_k \in C_f$ be a cell, then according to Def. 2.3 $cc_{fanin}(c_k) = deg_G^+(c)$ is the indegree, thus the *fan-in* of $c$ and $cc_{fanout}(c_k) = deg_G^-(c)$ the *fan-out* of $c$.

**Definition 5.2.4 Conditional Complexity.** Let $c_k \in C_f$ be a formula cell of a given spreadsheet, with $c = (ca{:}CA, v{:}V, f{:}(\ldots) \rightarrow V)$ and $f \neq \emptyset$. If $f$ contains a Boolean expression, the conditional complexity $cc_{conditional}(c_k) = nrOfConditions(c_k)$ is stipulated by the number of conditional decisions.

**Definition 5.2.5 Nesting Level.** Let $c_k \in C_f$ be a formula cell of a given spreadsheet, with $c = (ca{:}CA, v{:}V, f{:}(\ldots) \rightarrow V)$ and $f \neq \emptyset$. $f$ is a nested expression shaped $f = f_1 \circ f_2 \circ \ldots \circ f_{nl} = f_{nl}(f_{nl-1}(\ldots f_1(\ldots)))$, then $nl$ is the nesting level $cc_{nesting}(c_k) = nl$ of $c_k$.

The more references are bundled in one single cell, the more connected with the graph the cell is. The fan-out of the cell is therefore a very important metric for the analysis of change impacts. Since (not only) novice spreadsheet users often struggle with conditional results and the "=IF()"-function, Def. 5.2.5 provides a metric where subjective and objective complexity estimation may overlap. Nested "IF"-clauses are even more difficult to understand. This also holds for common nested expressions. Even if in some spreadsheet systems the string length of formulae is restricted to 255 characters the formulae can get very complex.

The specification of what is complex or even too complex is challenging and subject to further research and future studies. Nevertheless, let us suppose that such a threshold exists for the formula complexity metrics above, e.g. a threshold $t_{nesting}$ that defines the number of nesting levels above which it becomes difficult to understand formulae. Then it is possible to narrow the spreadsheet comprehension problem to "complex" spreadsheet cells.

**Definition 5.2.6 Complex Formula Cells.** Let $S = (V, E)$ be a spreadsheet program. The union of the following sets then is the set of complex formula cells.

$$\{c_k \in C_f \mid cc_{pathDep}(c_k) \geq t_{pathDep} \lor cc_{pathRef}(c_k) \geq t_{pathRef}\}$$
$$\{c_k \in C_f \mid cc_{spreading}(c_k) \geq t_{spreading}\}$$
$$\{c_k \in C_f \mid cc_{fanin}(c_k) \geq t_{fanin} \lor cc_{fanout}(c_k) \geq t_{fanout}\}$$
$$\{c_k \in C_f \mid cc_{conditional}(c_k) \geq t_{conditional}\}$$
$$\{c_k \in C_f \mid cc_{nesting}(c_k) \geq t_{nesting}\}$$

The set of complex spreadsheet cells defined in this manner may serve as reference point for different visualization approaches or as entry point for a comprehension approach.

From a different perspective, the cardinality and nature of this set might even serve as basis for reviews or as key argument for porting a spreadsheet application into a conventionally and professionally developed software solution.

### 5.3 Further Complexity Arguments

To conclude this section some metrics of simple binary nature (criteria) are mentioned as they highly influence the design complexity of a spreadsheet program. The following criteria should be born in mind:
- Are there pivot tables included?
- Is any a procedural extension (VBA, Python) included?
- Are external data sources included in the spreadsheet program?
- Are user-defined functions included?

The incorporation of these application features in a spreadsheet program makes it more fragile, vulnerable to broken links, or to misuse due to missing documentation.

### 6 RESULTS

Departing from the different characteristics of spreadsheet programs from conventional programs a set of visualization mechanisms and a set of metrics have been presented. As results of these deliberations, one should focus on the spreadsheet maintainer sitting in front of a sheet she or he has not developed her/himself or has developed it long time ago and thus forgotten most of the details that went into its original design.

In this case, the metrics discussed in section 5 will help to identify the critical spots in this spreadsheet program. Thus, the maintenance programmer could first look at some of those metrics (some of them might, due to the application domain, be evident from the outset) and decide on this basis, which visualization approach would best suit the overall comprehension task. E.g., if the spreading factor is high using the 3-D Workbook Visualizer might be the tool to be used first. Using this tool will show patterns that might be classified as "pearls-on-a-rope" or "quails with n tentacles". Seeing these patterns might immediately lead to different hypotheses. They are to be tested on the formula level in order to manifest themselves into a concrete conceptual model.

Having such a model on the macro level (workbook level) will allow a separation of concern, such that now individual worksheets are addressed. There, a broader host of metrics have been defined and consequently a broader strategy of visualising and analysing a given sheet exist. Finally, arriving at the formula level, visualizations of given formulae by looking at their fan-in and/or fan out will be helpful in some cases. In other cases, the remoteness of data transfer as expressed by calculation chains will lead to dissections of sheets into different slices, focussing on other comprehension aspects.